How Do Global Audiences Take Shape? The Role of Institutions and Culture in Patterns of Web Use

Harsh Taneja[1]

James Webster[2]

Suggested Citation:

**Acknowledgements**
The authors thank James Ettema, Edward Malthouse, Noshir Contractor, Robert Ackland, Angela Xiao Wu, Brian Keegan and Pablo Bockzkowski for their contributions to this research. We also thank Rufus Weston (then at BBC Global News) for facilitating access to comScore data. Finally we acknowledge the terrific editorial guidance by Silvio Waisbord.
This study is part of the first author's doctoral dissertation work and was partially supported by a Graduate Research Grant from The Graduate School at Northwestern University.

[1] School of Journalism, University of Missouri School of Journalism, Columbia, MO 65211

[2] Department of Communication Studies, Northwestern University, Evanston, IL, 60208

Corresponding Author: Harsh Taneja; email: harsh.taneja@gmail.com

**Abstract**

This study investigates the role of both cultural and technological factors in determining audience formation on a global scale. It integrates theories of media choice with theories of global cultural consumption and tests them by analyzing shared audience traffic between the world's 1000 most popular Websites. We find that language and geographic similarities are more powerful predictors of audience overlap than hyperlinks and genre similarity, highlighting the role of cultural structures in shaping global media use.

## How Do Global Audiences Take Shape? The Role of Institutions and Culture in Patterns of Web Use

Through the power of technology, age-old obstacles to human interaction, like geography, language and limited information, are falling and a new wave of human creativity and potential is rising (Schmidt & Cohen, 2013, p. 4).

This quote, from a book by the long-time CEO of Google, reflects an optimistic yet curiously deterministic prophecy about the revolutionary potential of the Internet. Such prophecies, common in popular discourse, predict that technologies and institutional structures will shape patterns of global cultural consumption, sweeping away old allegiances based on cultural traits such as language and geography. Scholars of global culture expect something quite different. They believe in the power of "cultural proximity," and the continued ability of cultural structures like language and geography to shape audiences (Consalvo, M. 2011). They argue that people prefer media closer to their own culture (Straubhaar, 1991 ) and, empirical studies show that as Internet use has deepened and broadened, users take on its topography heterogeneously (e.g., Burrell, 2012 Miller & Slater, 2000).

How does each of these factors shape global audiences? How relevant are cultural factors such as language and geography in determining global patterns of media consumption in an age when technologies and powerful institutions increasingly facilitate cross-border flow of content? Will technological infrastructures trump cultural differences or will they work in tandem to shape the patterns of global cultural consumption? This study is an attempt to answer those questions. It offers an empirical investigation of the role these factors play in determining audience formation on a global scale. It does so within a theoretical framework that integrates theories of media choice with the theories of global cultural consumption and extends extant empirical work on global Internet use with a large global sample of Websites, which includes a wide variety of Web domains.

The top 1000 Websites around the world account for about 99% of all traffic on the World Wide Web. We analyze data on audience duplication (the extent to which users who access site A also access site B) across these sites to identify patterns of global Web use. To explain the level of audience duplication, we use cultural factors such as similarity of language and geography, and institutional factors such as hyperlinks between each pair of Websites. We find that global Web usage largely clusters according to language and geography of these Websites, and not according to their content genres. Further, we find a very low correlation between hyperlinks and audience traffic, suggesting that hyperlinks are not as powerful a determinant of Web use as they are commonly thought to be. These findings contribute to existing research on media choice as well as the literature on global media flows.

**Theory**

We draw on two distinct bodies of theory to frame this research. The first is the literature on media choice. In particular, we make use of newer "integrated" theories of choice. These consider the roles of both individual predispositions and structures in shaping media use. The second is the literature on media globalization. Broadly speaking, it adopts one of the two theoretical frameworks to explain global cultural consumption: one privileges institutional structures like hyperlink architecture; the other emphasizes cultural structures like language. We recast the literature on media globalization into the integrated theories of media choice, and hypothesize that audiences on the World Wide Web enact their preferences within both institutional and cultural structures when consuming online content.

**Integrated theories of media choice**. Historically, theorists have relied heavily on individual predispositions (e.g., attitudes, tastes, gratifications sought, program types preferences, etc.) to explain media choice. Most of these expect audiences to evidence some sort of content loyalties. The abundance of options now available to audiences has given rise to a pervasive rhetoric of individual empowerment (e.g., Jenkins, Ford, & Green, 2013; Napoli, 2011). In theory, this should manifest itself in genre loyalties reflecting individual tastes or attitudes or program type preferences. Hence, we might expect that fans of news would consume all of the best available news sites, or that partisans would be attuned to ideologically consistent sites no matter their geographic origin, and so forth. In a world of anytime, anywhere media, it is especially tempting to think that individual preferences are all that matter in determining audience behavior. But structural factors have a powerful influence on patterns of media use in the new digital environment (Webster, 2014). Recent studies find that social and technical structures, such as routines and access to platforms, are more influential than individual traits in explaining television viewing and digital media consumption (Cooper & Tang, 2009; Perusko, Vozab, & Čuvalo, 2015; Taneja, Webster, Malthouse, & Ksiazek, 2012; Webster & Ksiazek, 2012; Wonneberger, Schoenbach, & van Meurs, 2009, 2011).

Integrated theories that consider both individual and structural factors, are sometimes grounded in a structurational framework (Giddens, 1984; Webster, 2011). In a nutshell, structuration sees agents (media users) drawing on the resources of the media to achieve their own ends. These resources include the available technologies, programs and services. As agents use media, they reproduce and alter the structural features of the environment. In this view, agency and structure are mutually constituted, something Giddens called a "duality" (1984, p. 25).

Despite its conceptual elegance, structuration has only been used to study audience behavior in single markets or to make limited cross-country comparisons (e.g., Perusko, et al., 2015; Yuan & Ksiazek, 2011). However, we believe that this theoretical framework can be especially useful in studying audience formation on a global scale where one would expect to find greater variation in structural factors. In particular, our study focuses on understanding how a variety of technological and social structures shape large-scale patterns of traffic across Websites. What follows is a brief review of the theories that explain the flow and consumption of media products across national boundaries.

**Theories of global media consumption.**

Cultural objects, including images, languages, and hairstyles, now move ever more swiftly across regional and national boundaries. The acceleration is a consequence of the speed and spread of the Internet and the simultaneous comparative growth in travel, cross-cultural media and global advertisement (Appadurai, 2013, p. 61).

Cultural products were exchanged between civilizations even in ancient times, often facilitated by the movement of people. In the last two decades of the twentieth century, the growths of satellite and cable television followed by the Internet have made such flows nearly ubiquitous. So, how do audiences make choices between media products from abroad and those from their home country? Broadly, there are two schools of thought.

Illustrative of the first school is the thesis of "cultural imperialism" (Schiller, 1969), which states that global media flows are heavily imbalanced, with the majority of audiovisual content exported by the developed world (mainly the US) to the developing world. Such imbalances have given rise to fears that global cultural consumption will become homogenized. Neoclassical economists have also noted this imbalance in global flows of media and attribute this to the superior production values and distribution muscle of large and wealthy countries. These arguments are buttressed by newer studies of Internet structure, often mapping the WWW using hyperlink analysis, which find that the Western world (especially the US) is at the center of global information flows (e.g., Park, Barnett & Chung, 2011; Barnett & Park, 2014).

In opposition, the cultural proximity thesis asserts that given a choice, audiences around the world will tend to choose culturally proximate content (e.g., Straubhaar, 1991) and that foreign products are more successful if the cultural distance between importing and exporting countries is low (Fu & Govindarau, 2010). Moreover, audiences across cultures interpret the same content in very different ways (Liebes & Katz, 1990), which is a reflection on their own cultural identity (Kraidy, 2002). These predispositions shape the global media landscape as a mosaic of "culturally defined markets" along geo-linguistic lines, based on factors such as shared geographies, languages, colonial history, ethnicity, etc. (Straubhaar, 2007; Sinclair, 1999).

Integrated theories of media choice offer a way to test these seemingly opposing bodies of literature. Essentially, the "imperialism school" privileges the role of institutions such as media industries and government regulators in structuring global cultural consumption. In the case of the Internet these structures manifest both as enablers, such as hyperlinks[i] and search engines, and censors such as Internet firewalls. Henceforth, we'll call these "institutional structures." On the other hand, the "cultural proximity" thesis posits that cultural factors, such as the languages people speak or their national and regional identities, are more powerful drivers of media consumption. Henceforth, we refer to these as "cultural structures."

We argue that both types of structures, institutional and cultural, should contribute to audience formation on a global scale, although the explanatory power of each remains to be determined. We also believe that studying people's use of the World Wide Web offers a way to test how each set of structures contributes to global cultural consumption.

**Global cultural consumption on the World Wide Web**. It is not just the sheer size but also the structure of the World Wide Web that makes it a truly global medium. The Web is essentially billions of documents (or pages) that are connected to each other by hyperlinks. Studies of hyperlinked structures reveal that even as the Web has grown, its diameter or the average degrees of separation between pages has remained relatively small (Barabási, 2009). These features suggest that Internet users can easily access content anywhere on the Web with the use of hyperlinks, so long as it is not censored or concealed behind pay walls. Search engines make it even easier for people to access whatever is out there. Further, automatic machine translation tools such as *Google Translate* allow people to consume foreign language content in a language of their choice. These features have tempted many scholars and public intellectuals to envision a completely connected, globalized world of Web users. Relatively few, however, have noted that audiences continue to pay "disproportionate attention to phenomena that unfold nearby and directly affect ourselves, our friends and our families" (Zuckerman, 2013, p. 19). For instance, people form ties on Twitter with users in close geographic proximity or who share the same language, or when there is direct air connectivity between their locations (Takhteyev, Gruzd, & Wellman, 2012). Likewise, people are more likely to email someone who belongs to the same culture (State, Park, Weber, Mejova, & Macy, 2015). Even Wikipedia, the so-called global online encyclopedia, is quite "local" if one closely analyzes differences between its many language versions (Hecht & Gergle, 2010).

It is clear that, faced with an abundance of Websites, people wouldn't access all available content but create repertoires containing a small number of Websites they visit regularly. What is not as clear is the basis on which they would create these repertoires. More specifically, are these repertoires likely to be based on genre preferences, without regard to cultural affinities or are they driven by cultural proximity? Our theories of global consumption suggest two possibilities.

First, since users who have access are free to consume what they want from any part of the world, we may indeed find that audiences gravitate towards the "best" content of a type. For instance, someone who is fluent in English and interested in news and current affairs could go to the *New York Times* or the *Guardian* provided this user considered these the best sources. If this were the case, online content produced in the Western markets that have traditionally dominated much of global media marketplace would be more popular than locally produced content. This outcome would be consistent with the picture painted by the structure of hyperlinks, where sites from most countries appear to link generally to Websites from the US and Western world (Barnett & Park, 2014) and "the economy rather than culture is the primary determinant of the structure of international hyperlink flows" (Barnett & Sung, 2005, p. 230).

Alternatively, provided enough local content is available online, audiences are likely to consume culturally proximate Websites, ones that are in languages they prefer. They need not pay attention to Websites that they don't consider relevant or that are in foreign languages. In fact the relative ease of access to content online compared to TV and films is likely to enhance this tendency. Besides, barring a few expectations, institutional regulators do not control the supply of content online to the extent they do in traditional media. These distinct possibilities motivate the following, rather broad, research question.

>RQ1. What patterns of consumption are evident when audiences around the world navigate on the World Wide Web?

We know that, with a few exceptions, audiences in any part of the world have access to both domestic and foreign products on the WWW. In such a scenario one may expect most people to consume both domestic and foreign products, in varying proportions. Therefore, to capture all such behaviors in the aggregate, we need to analyze audiences to a media outlet in relation to all the other media outlets they use. One way to do that is through the analyses of "audience duplication," or the extent of audience overlap between pairs of media outlets (such as Websites.) Audience duplication, measured across all pairs of media outlets, can easily be conceptualized in network analytic terms, with media outlets being the nodes connected to each other based on the extent of audience duplication (see Webster & Ksiazek, 2012; Taneja &Wu, 2014).

First, it seems likely that audiences prefer to consume content in a language they are most comfortable with. Therefore, audiences who visit a Website in a given language would also visit other Websites that are available in the same language. Hence, we expect Websites in the same language to have higher duplicated audiences than Websites in different languages. Other than language, it may be that audiences prefer content that focuses on their own geographies (countries). That is, when domestic (national) content is available audiences will prefer it to content from abroad. Therefore, audiences who visit a Website that focuses on a country would most likely access other Websites that also focus on the same country

In most cases, countries have a majority language. However, many countries share languages, as well as audiences within a country can speak different languages. Hence, we would expect audience duplication between two Websites to be higher when they both share the same language and focus on the same geography than when they share either language or geographical focus alone. As an example, the duplication between two Swiss German sites would be higher than duplication between two Swiss sites that are in French and German languages. Further, duplication would also be higher than that between a German site that focuses on Switzerland and a German site that focuses on Germany. This leads to the following hypothesis.

> H1: Similarity of geographical focus combined with language similarity results in higher audience duplication between Websites than language similarity or geographical focus alone.

Second, since hyperlinks are thought to help users navigate between two Websites, it is conceivable that they impact the extent of audience duplication between them. Studies of global hyperlink structures show that, although most language sites generally link to sites in the same language, sites in English tend to be more central than sites in other languages (Daniel & Josh, 2011). The same is true for developed countries in comparison to the developing World (e.g., Barnett & Park, 2014). But it is unclear whether people actually follow those links. That is, to what extent will non-English speakers click on the many hyperlinks to English Websites? A few recent studies using data from Alexa.com have correlated hyperlink structure with shared Website usage between countries (Barnett & Park, 2014) or with clickstreams (Wu & Ackland, 2014) and these show that usage and hyperlink networks need not correlate. This motivates the following research question:

> RQ2: Do hyperlinks between Websites explain (and if so to what extent) the level of audience duplication between them?

Finally, genres have been a popular way to explain audience loyalties in older media. Given the relative ease of access on the Web, it is conceivable that certain types of Websites may be more popular across linguistic and geographic regions, whereas certain others may be more restricted in their appeal, as they focus on a specific geography and/or have content in one language. These possibilities motivate the final research question:

> RQ3: Does genre similarity between Websites explain audience duplication between them?

**Method**

    **Data.** We obtained the "audience network" using data from a global Internet audience measurement panel described below. In addition, we "crawled" the same set of domains for which we had Web traffic data, to obtain a parallel network, where we considered individual Websites as nodes tied to one another on the basis of hyperlinks between them. The latter, we refer to as the "hyperlink network."

    *Audience Network.* We used data from comScore[ii], a panel based service that provides Internet audience measurement data once a month. With approximately 2 million panelists in 170 countries under continuous measurement, the comScore panel utilizes a meter that captures behavioral information through a panelist's computer. Data are collected from both work and home computers of the panel members, who are recruited based on a telephonic enumeration survey to establish the Internet universe in each country. comScore organizes Websites by Web domains and subdomains. Our sample included the top 1000 Web domains (ranked by monthly unique users). This sample accounts for 99% of Web user visits, and ensures an adequate representation of sites in different languages and different geographies. For many large Websites such as Google, the different geo-linguistic variants are classified as separate domains (e.g., www.google.es, www.google.de etc.). For certain large domains such as Wikipedia, language versions are sub-domains of the main domain (e.g., es.wikipedia.org). In such cases, these sub-domains have been considered in the final sample instead of restricting to Web domains. These

data reflect traffic during June 2012, and 973 Websites of the top 1000 were included in the final sample. Together these sites provided content in 50 languages (many sites were in multiple languages) and between them focused on 43 countries (some sites had an explicit global focus, see next section). For each one of the 973 Websites, we obtained its audience duplication with all other 972 sites. Thus the final dataset has 472,878 (= (973 *972)/2) pairs of audience duplication numbers.

**FIGURES 1 AND 2 ABOUT HERE**

*Hyperlinks Network.* In addition to the traffic data, we obtained data on hyperlinks (during August 2012) between Websites using a crawler called VOSON. This crawler allows the user to specify all the Website addresses for which one needs information on inbound hyperlinks and downloads them systematically. Later it analyzes all the links present on the downloaded pages to provide hyperlinks between all pairs of initially specified Websites. Further, for this study, the creators of VOSON also queried a search engine, Blekko, to provide indexed information on incoming links to Websites. Popular search engines such as Google and Bing no longer provide an API for this. This helps capture any links missed in the original downloads done by VOSON. This resulted in a matrix that contained the number of hyperlinks between all possible pairs of 961 sites drawn from the same sample of 1000 most popular domains as the audience duplication. Some sites do not allow crawling and hence had to be dropped. The two matrices provided largely comparable data on hyperlinks and audience duplication between the same pair of most popular Websites. 950 domains in both networks were identical and hence yielded networks with identical nodes.

**Analysis.**

***Descriptive Network Analysis.*** The nodes in our audience network are Web Domains (such as www.google.com and www.google.co.uk, two separate Web domains), which are connected to each other based on audience duplication. However, popular domains are likely to have audiences in common due to random chance. Therefore, in order to declare a tie, we have set the bar above the level of duplication expected by chance alone. Ksiazek (2011) describes this procedure. In essence, if the observed level of shared audiences (observed frequency) was greater than the expected level based on random chance, we regarded the two outlets as tied. Conversely, if the level of observed duplication was less than or equal to that expected by random chance, we disregarded the tie. This procedure is quite similar to how residuals are calculated in chi-square analysis.

To identify and describe the resulting patterns from the ways in which Web audiences navigate around the World Wide Web, we perform descriptive network analysis of the "audience network" using *dichotomous ties*. Specifically, two measures, *network centralization* and *clustering coefficient*, which indicate the overall shape of the network, answer RQ1. *Network centralization* indicates to what extents are the ties concentrated on a small set of nodes or more uniformly distributed. A high centralization score would indicate that a few nodes receive the lion's share of ties and these would be highly central in the network. In a network with low centralization score, however, the ties are more evenly distributed, and it is hard to identify a set of few central nodes that receive most of the links. Network centralization for a media audience network such as this one can be considered analogous to the Herfindhal-Hirschman index, a well-established measure of market concentration (Webster & Ksiazek, 2012). Likewise *clustering coefficient* indicates if nodes in a graph tend to cluster together. This is calculated by considering sets of three nodes or triplets that occur in the network. Any triplet of nodes can form either three open triangles or one closed triangle. The clustering coefficient for each node is the ratio of the number of closed triangles that exist in the network to the total number of triangles (both closed and open) theoretically possible (Watts & Strogatz, 1998). It varies between 0 and 1, with 1 indicating a fully clustered network and 0 indicating no clustering.

***Regression Analysis.*** Although descriptive network analysis and identification of clusters does provide a detailed description of patterns of audience flow, it does not help explain the variance in audience duplication. Therefore to confirm Hypotheses H1 as well as to answer RQ2 and RQ3, we employ multiple linear regression analysis. The dependent variable is the audience duplication between sites. The independent variables are language similarity, geographic similarity, genre similarity and number of hyperlinks between all pairs of Websites. As already noted, for 950 sites (out of 973), we had data both on audience duplication with all other sites, as well as hyperlinks between all possible pairs[iii][iv]. We use these 950 sites in the regression model.

*Dependent Variable.* *Audience Duplication:* For the 950 sites we include in the regression models, which amounts to 450,775 (approx. one-half million) duplicated pairs. We take the natural logarithm of this variable to symmetrize its right skewed distribution, as is a common practice with amount and count variables.

*Independent Variables. Language Similarity:* We use Jaccard's similarity coefficient to compute the extent of language similarity between Websites (Gower, 1985). The Jaccard coefficient for any pair of Websites is the ratio of languages in common between the pair to the total number of languages between them. For instance, if site A is in English and Spanish and site B is in Spanish and French, the Jaccard coefficient would be 1/3 since Spanish is the common language and these Websites between them have three languages in all (English, Spanish and French).

*Same Geographical Focus:* We identified the focal geography of every Website in the sample. In the majority of the cases (2 out of 3) they could be assigned to a single country. Triangulating various sources of information arrived at this assignment? First, many Websites have country-specific domain names, which signal their foci on a specific country. Second, the language of a Website is often a good indicator of the geography it focuses on. For instance, a site in Russian focuses on Russia or former Soviet countries such as Ukraine and sites exclusively in Spanish focus on Spain or Latin American countries. Third, we studied the "about us" pages to read the site's mission statements where they often declare their geographical focus. Finally, we relied on third party sources such as Alexa.com (a Web information company) that maintains ownership, content, marketing and traffic information about most Websites. Overall, we were able to assign a focal geography to about 650 sites. The remaining 300 sites had versions in multiple languages that catered to multiple geographies. We coded such sites as "Global". Examples of such sites would be platforms that rely on user-generated content such as Youtube.com, Facebook.com, Blogger.com and Twitter.com and corporate Websites of multinational corporations that had multiple languages/countries such as Microsoft.com and Nokia.com. To use in the regression model, we coded each pair of sites that focused on the same country, or with at least one "global" site as '1'. All other pairs were coded '0'.

*Same Genre:* In addition to language and geography, we categorized each Website by its content genre. There is no universally accepted categorization scheme in the literature and hence we relied on the genre categorization comScore uses in its traffic reports. comScore uses a total of 27 basic genre categories and further categorizes sites into subgenres within each of these genres. For the regression analysis, we coded the genre similarity for each pair of Websites as '1' if they belonged to the same genre or '0' if they were assigned a different genre by comScore.

One can quibble with the way comScore categorizes sites, but the most important consideration in this analysis is whether or not the two sites belong to the same genre.

*Hyperlinks:* Using the results returned by VOSON, we computed the total number of links that existed between any pair of Websites as the sum of links in either direction. In other words, for any Website pair, say, A and B, we considered the number of hyperlinks between A and B as the sum of the number of hyperlinks from B to A as well as those from A to B. This being a count variable, we took a natural logarithm to symmetrize its distribution.

*Audience Size:* To control for the effect of size (as popular Websites may have high audience duplication with other sites), we include the size of either outlet in each Website pair in the regression model. We operationalized this variable as the product of the size of the two outlets and took its logarithm to symmetrize the distribution. This method of handling data on audience duplication is described in Webster (2006).

**Results**

    **The Shape of Online Audience flows.** First, we use some global measures (for the entire network) to describe the shape of audience flows. The network centralization score for the audience network is 52%, a moderately high score (centralization ranges between 0% and 100% where 100% indicates a perfect star structure with all nodes connected to one central node and not to any other node). In terms of audience flows, this suggests that there are a relatively small number of sites that do get links from most sites and these are quite central to the network. This is not surprising given that the top Websites (even within the relatively small sample of 973) get a disproportionately high amount of traffic and, consequently, have audience overlaps with large number of sites.

Network centralization alone does not reveal the complete shape of the network. As noted, the clustering coefficient indicates the average tendency of any three nodes in the graph to form a triangle (i.e., a connected triad). The local clustering coefficient for the audience network is .846 (weighted .752), which is a very high score (clustering coefficient varies between 0 and 1 and 1 indicates a completely clustered network.) This high clustering coefficient suggests that Websites cluster into groups in a manner where all sites belonging to the same group have high audience duplication between them and Websites belonging to different groups have relatively low duplication between them.

It is important to consider the high values of clustering coefficient and the network centralization together. In combination they suggest that while the network ties are concentrated on a small number of nodes (as suggested by the high centralization scores), the network has a high tendency to break into subgroups. These highly central nodes (with the largest share of ties) are most likely distributed across these clusters. Thus, the overall audience flows are quite concentrated, but all concentrated nodes probably lie in separate clusters within this network. This is confirmed by a visual inspection of the network diagram shown in Fig 3.

**FIGURE 3 ABOUT HERE**

Figure 3 shows a visualization of the audience network. The dots are the nodes (Websites) and the lines the ties (based on audience duplication) between them. This visualization based on the algorithm of Fruchterman and Reingold (1991) belongs to a class of visualization techniques known in graph theory as force directed graphs. The basic mechanism is that there are repulsive forces between all nodes; however, nodes that are adjacent to each other also have attractive forces. Corresponding to this description, in the final visualization, groups of nodes can be seen adjacent to each other (i.e., tend to have ties with one another) group to form tightly knit clusters with relative separation from other groups of nodes that are adjacent to each other. Broadly speaking, the visualization explains that global usage of Websites is clustered according to geography and language. The regression models, reported as follows, confirm this.

**Explaining Global Online Audience Flows.** The second component of the analysis is aimed at testing the power of both cultural structures as well as institutional structures in an integrated model to explain global online audience flows. We know from literature as well as the analysis above that both cultural factors such as language, focal geography and institutional factors such as hyperlinks could explain the patterns we just described in section 1. The regression analysis tests H1, and answers RQ2 and RQ3.

To recap, the dependent variable in the regression analysis is the audience duplication between all possible pairs of Websites (obtained from the audience network). The independent variables are *language similarity*, *geographic similarity, genre similarity* and *number of hyperlinks* between each of these pairs of Websites. Our sample is the 950 domains common to the hyperlinks network and the audience network, resulting in $N = 450{,}775$ cases with no missing values. We logged all amount/count variables to symmetrize their distributions.

Further, while specific countries generally speak a specific language, often a given language is spoken in more than one country. Therefore, Website pairs that share both language and a geographical focus may be likely to have more audience overlap than Website pairs that share either language or geographical focus alone. Hence, to test this *difference*, we included an interaction term of *language similarity * geographic similarity* in the model, which tests H1.

Before reporting the regression models, we report pairwise correlations between all pairs of independent variables (Table 1). The table clearly reveals that are no large correlations between any of the predictors that can pose issues of multicollinearity or of masking the total amount of variance explained by each.

**TABLES 1 AND 2 ABOUT HERE**

In Table 2(a), we report the results of the models. Instead of including all variables in one go, we decided to include each variable one by one, in the order of our theoretical build up, and build a series of models, in each of which we control for the fixed effects of Website size. Consequently, the first model (column 1), we include only the language similarity as the independent variable along with controlling for size. Unsurprisingly, language similarity has a large and significant positive coefficient. Next, we introduce geographic similarity and although the magnitude of the language similarity coefficient reduces, it still remains large and highly significant. Geographic similarity, too, has a large and positive significant coefficient. In the next two models (models (3) and (4)) we introduce genre similarity followed by hyperlinks. They have significant positive coefficients but these do little to reduce the magnitude or significance of the coefficients associated with either language or geography. Language and geography each have a significant as well as sizeable effect.

Finally, to test Hypothesis H1, we included an interaction term as a product of language and geography. In itself it has a significant negative coefficient, suggesting that having the same geography reduces the impact of language alone on levels of audience duplication. Hence, here we computed the net effect (including the main and interaction effect) for language by varying the levels of geography. In the first case, we consider when the two Websites have the same geographical focus, and hence the term 'geographic similarity' in the equation takes the value '1.' Therefore, the net effect of language is **β** (language) + **β** (language * geography) = 2.882 (4.662- 1.78* 1). However, when the geography is dissimilar (i.e., it takes the value 0), the net effect of language will be **β** (language) + **β** (language * geography) = 4.662 (4.662 -1.78*0).

Corresponding to the explanation offered above, in order to compute the effect of the regressor on the dependent variable, we estimated the values of the dependent variable for various levels of language similarity using two specifications: one when the two sites have matched geographies (geographic similarity = 1), and another when they focus on dissimilar geographies (geographic similarity = 0). In doing so, we held all other variables at their mean values. Figure 4 shows the result. It is evident that for all levels of language similarity, focusing on the same geography makes it more likely for two Websites to have duplicated audiences than focusing on dissimilar geographies. Hence, H1 is supported.

**FIGURE 4 ABOUT HERE**

As a final step, we report the standardized coefficients corresponding to the final model in Table 2(b). Since the independent variables are in different units and some are logged, whereas others are not, standardized coefficients provide a comparable metric to examine the effect sizes of different variables. These basically represent the change in the dependent variable due to a change in one standard deviation of the independent variable. These again confirm that similarities in language and geographical focus have much greater effects than either number of hyperlinks or genre similarity alone.

**Discussion**

The technologies used to create, store and distribute media have made the content produced in one country readily available to people in other countries. While this has been true of films and television for some time, the World Wide Web has dramatically increased the amount of cross border content available to anyone with Internet access. The options available to the average Internet user are mind-boggling. However, there has been real debate about what factors will govern the choices people make. Will high production values and technical advantages win users over, or will the drawing power of cultural proximity trump those institutional structures? This study contributes to that debate by explaining how Web audiences around the world make choices between domestic and foreign content. In doing so, it is one of the first studies to test the relative power of both cultural and institutional structures. It also demonstrates how structuration can be used to study global audience formation, an important contribution to the literature on media choice. Further, the study enhances our understanding of cultural proximity by parsing the relative contribution of geography versus language (two well recognized but often conflated factors) in shaping global Web use.

Our analysis provides convincing evidence that global audiences cluster based on language and geography. Specifically, we find that similarity of languages *and* a common geographical focus of any two Websites offer the best explanations of audience overlap between sites. Conversely, we find the number of hyperlinks between Websites explains very little audience overlap, suggesting that Web users don't necessarily follow hyperlinks. In other words, cultural structures like language and geography are far more potent determinants of global audience formation than institutional structures like the presence of hyperlinks or firewalls. Below, we discuss each of these findings in some detail with an eye toward their theoretical contributions. Finally, we discuss the larger implications of this study and speculate on the future of global cultural consumption.

**Parsing the Contributions of Language and Geography.** Consistent with the cultural proximity thesis, this study finds language to be a major factor associated with well-defined communities or clusters of Websites consumed by the same set of users. While research has shown that "minority language" offerings within larger countries spawn "small but loyal" audiences (e.g., Ksiazek &Webster, 2008), we find language to be an instrumental force in creating communities of "large and loyal" audiences. In other words, users who consume Websites in one language, will consume other Websites in the same language, and are less likely to consume Websites outside that language. Hence, language acts as glue that binds WWW users who speak the same language into a large community.

However, geography is clearly confounded with the role that language plays in explaining audience formation. Often people in geographically contiguous regions speak the same language and that makes it difficult to isolate which of these two factors is more effective in creating an affinity for local or culturally proximate content. Our research takes advantage of very large datasets to disentangle these factors, and in doing so, adds to the literature on cultural proximity, which has *separately* highlighted the importance of language or geography in audience formation (e.g., Straubhaar, 2007). The regression model used in this study, in particular the interaction term between language and geography, helps isolate the role of each. It clearly suggests (see Figure 4) that when two sites are similar in language *and* geographical focus, they will have more duplicated audiences than when they have language similarity or geographic similarity alone. The result indicates that while language is an enabler, a tool that facilitates audiences to consume content, it is the relevance of the content that matters more. And relevance is often a matter of geography. As Ethan Zuckerman noted,

> It makes sense that linguistically isolated nations – nations that don't share a language with any other countries – like Japan or South Korea would read few international sources. It's more surprising that despite a long colonial legacy, and shared language, Indians and Brits don't read more of each other's content. Nor do they visit US news sites. As it turns out shared language offers no guarantee of interest in each other's media (2013, p. 59).

**Why not Genres?** The study finds that "genre similarity" between Websites has a rather small, almost negligible, effect on audience duplication. This is a curious finding. There is very little evidence of these individual appetites and predispositions in our results. It could be that our data, which does not allow us to track individual users or their declared preferences, are too insensitive to individual differences to detect genre loyalties. It could also be that our genre classifications are too broad to reveal subtle but meaningful patterns of audience loyalty. Given the theoretical importance of genre preferences in media research, testing a more nuanced categorization scheme may worth the effort. But if we take the results at face value – that genre effects are basically non-existent – we should also consider why.

The first and most obvious explanation is that despite our theoretical expectations, genre preferences are not all that powerful. There is some evidence that genre preferences, far from being fixed, are constructed on the fly or that people are simply "omnivorous" in their preferences. Second, it may also be that the sheer abundance of choices creates problems of "bounded rationality," making it hard for people to connect their desires with their actions. Third, even if we assume that clear, enduring, genre preferences always guide media choice, it may be that one Website of a sort is enough to gratify whatever needs are being acted out (see Webster, 2014 for a discussion of these factors). For now, it appears that people seek variety in their daily media diets and that users who speak a particular language are accessing all manner of content in the language that focuses on their geography. Broadly speaking, this finding highlights that social structures moderate the role of individual predispositions and contributes to the

HOW DO GLOBAL AUDIENCES TAKE SHAPE? 25literature on how structures shape audience choices in digital media (e.g., Taneja et al. 2012, State et al, 2015)

**Why Don't Audiences Follow Hyperlinks?** This study confirms that WWW structure based on hyperlinks is not indicative of the navigation patterns of online audiences, measured through pair wise audience duplication between Websites. In doing so, it lends external validly to recent studies that found various measures of usage such as clickstream (Wu & Ackland, 2014) and site popularity in a country (Barnett & Park, 2014) to be uncorrelated with hyperlink analysis. This lack of correlation between hyperlinks and audience highlights a glaring limitation of hyperlink analysis: In counting links alone, researchers have very limited information about the inherent motivations of the link provider, and these may differ across Websites. The simplest dichotomy being that a link could signal either praise or criticism (De Maeyer, 2013). Further, the intention of the link provider may be contained in either the link itself or in the linked resource. Finally, Websites in different genres may have different linking strategies. A news Website may link to an influential Website to signal legitimacy whereas a personal homepage may have links representing pages and organizations personally relevant to the author. In our dataset we find a large number of hyperlinks between the Spanish and English subdomains of CNN Websites, but little audience overlap between two. Likewise, Wikipedia generally has hyperlinks between all articles that cover the same concept across languages, however people on average don't appear to access the same information in multiple languages.

Hence, the differing purposes and motivations for providing hyperlinks make it difficult to say with any reliability whether the structure of links would predict the structure of online traffic patterns. Alternatively, this study highlights the value of using audience duplication as an alternate analytical tool in conceptualizing the WWW structure. Such a conceptualization has the potential to explain mechanisms driving behavior of online audiences that hyperlink analysis cannot reveal. It allows for examining the role of hyperlinks alongside other site-specific attributes such as language, geography and genre.

**The Future of Global Cultural Consumption**

We conclude by speculating on factors, which will shape patterns of global media use, as digital media become even more ubiquitous and the global communication infrastructures become even more interconnected. Many continue to believe that institutional factors will rule the day.

> …[I]nformation flowing from the global north to the global south will have the greatest impact on converging values in cosmopolitan societies characterized by integration into world markets, freedom of the press, and widespread access to media. Parochial societies lacking these conditions are less likely to be affected by these developments (Norris & Inglehart, 2009).

We argue that both institutional and cultural structures will work in tandem to shape global cultural consumption. Norris and Inglehart (2009), like Schmidt and Cohen (2013), seem to believe that global cultural diversity will persist because these institutional factors, or "firewalls" prevent *parochial* nations from becoming *cosmopolitan*. In doing so, they seem to ignore the power of cultural factors in shaping patterns of global cultural consumption. This study suggests otherwise.

Based on the findings here, it appears that cultural factors, such as language and geography, are extremely powerful forces in shaping global patterns of cultural consumption. And these factors are at work irrespective of the nature of these so-called institutional firewalls. For instance, consistent with Taneja and Wu (2014), our visualization also suggests that Chinese sites are no more or less "isolated" into a cluster of their own as the ones from Japan. The former has a huge Internet firewall; the latter is an exemplar of an open democracy. In other words, restrictions imposed by institutions are not the only barriers that will help maintain culturally proximate consumption. Conversely, removal of all institutional restrictions and achieving a completely connected world where information can flow freely will not homogenize cultural consumption, either. The relative roles of different factors may shift, and so are worth tracking in subsequent research.

Ideally, future studies will be able to exploit individual level data. This would give investigators a clearer view of agents and how they operate within the institutional and social structures that surround them. It would also allow for a more complete test of fully integrated models such as those developed and tested for television viewing (e.g., Cooper, 1993; Taneja & Viswanthan, 2014; Webster & Wakshlag, 1983; Wonnebeger et al 2009). Our best chance of understanding global cultural consumption will come only when we include the full array of individual and structural factors that shape our actions.

HOW DO GLOBAL AUDIENCES TAKE SHAPE?                                                                33Yuan, E. J., & Ksiazek, T. B. (2011). The duality of structure in China's national television

    market: A network analysis of audience behavior. *Journal of Broadcasting & Electronic

    Media, 55*(2), 180-197. doi:10.1080/08838151.2011.570825

Zuckerman, E. (2013). *Rewire* : *digital cosmopolitans in the age of connection*. New York:

    Norton and Company**Endnotes**

[i] Hyperlinks, although a technical feature, are often used by media organizations as a strategy to drive traffic to their own and affiliated online properties. Such strategies have been effective for many media organizations and serve as proxies for "partnerships and alliances between organizations, and even international flow of information" (Weber, Chung & Park, 2012 p. 117)

[ii] This information has been taken from comScore's own documentation on methodology.

[iii] It is conceivable that these pair-wise duplication values may not be completely independent. Violating this assumption of independence of observations could alter the standard errors, affecting the p-values, anyways likely to be significant here given the large number of cases. This wouldn't alter the effect sizes, as measured by the magnitude of the Beta coefficients, which our analysis anyway focuses on.

[iv] An alternative to regression is to use "exponential random graph modeling" that predicts whether an observed network tie has a greater or lower chance of forming compared to that tie forming in a random graph. This helps explain the presence of many endogenous network effects, which are conceived as an outcome of network structures rather than the node attributes. However in the present case, most anticipated structural effects can be explained by employing node attributes, some of which may not have been included in the study.

**Table 1:** *Correlation Matrix between Independent Variables*

|  | Language Similarity | Geographic Similarity | Genre Similarity | Hyperlinks (Log) | Size(Log) |
|---|---|---|---|---|---|
| Language Similarity | 1 |  |  |  |  |
| Geographic Similarity | .325** | 1 |  |  |  |
| Genre Similarity | .008** | .016** | 1 |  |  |
| Hyperlinks (Log) | .137** | .106** | .033* | 1 |  |
| Size (Log) | -.025** | .055** | -.001 | .194** | 1 |

** correlation significant at p <0.01

34**Table 2 (a):** *OLS Regression Models to Explain Website Audience Duplication*

|  | Beta (1) | T | Beta (2) | T | Beta (3) | T | Beta (4) | T | Beta (5) | T |
|---|---|---|---|---|---|---|---|---|---|---|
| Language Similarity | 4.136 | 344.0 | 3.248 | -214.7 | 3.247 | 261 | 3.187 | 255.5 | 4.662 | 156.6 |
| Geographic Similarity |  |  | 2.573 | 261.5 | 2.570 | 274.5 | 2.551 | 272.5 | 2.756 | 273.9 |
| Genre Similarity |  |  |  |  | 0.358 | 22.3 | 0.336 | 21.0 | 0.346 | 21.7 |
| Hyperlinks (Log) |  |  |  |  |  |  | 0.479 | 39.9 | 0.484 | .40.4 |
| Language * Geography |  |  |  |  |  |  |  |  | -1.78 | -54.5 |
| **Control** |  |  |  |  |  |  |  |  |  |  |
| Size (Log) | 1.329 | 314.8 | 1.258 | 321.0 | 1.258 | 321.3 | 1.227 | 307.8 | 1.224 | 308.2 |
| Intercept | -16.287 | -199.6 | -16.213 | -214.7 | -16.243 | -215.2 | -15.652 | -203.8 | -15.717 | -205.2 |
| ***R-Square*** | **.33** |  | **.42** |  | **.42** |  | **.42** |  | **.42** |  |

*N=450775. The dependent variable is the log of audience duplication between all website pairs.*

**Table 2(b):** *Standardized Coefficients corresponding to model (5) in Table 2(a)*

| Language Similarity | Geographic Similarity | Genre Similarity | Hyperlinks (Log) | Language * Geography | Size (Log) |
|---|---|---|---|---|---|
| 0.452 | 0.354 | 0.025 | 0.047 | -0.173 | 0.356 |

*Figure 1: Languages in the Sample.*

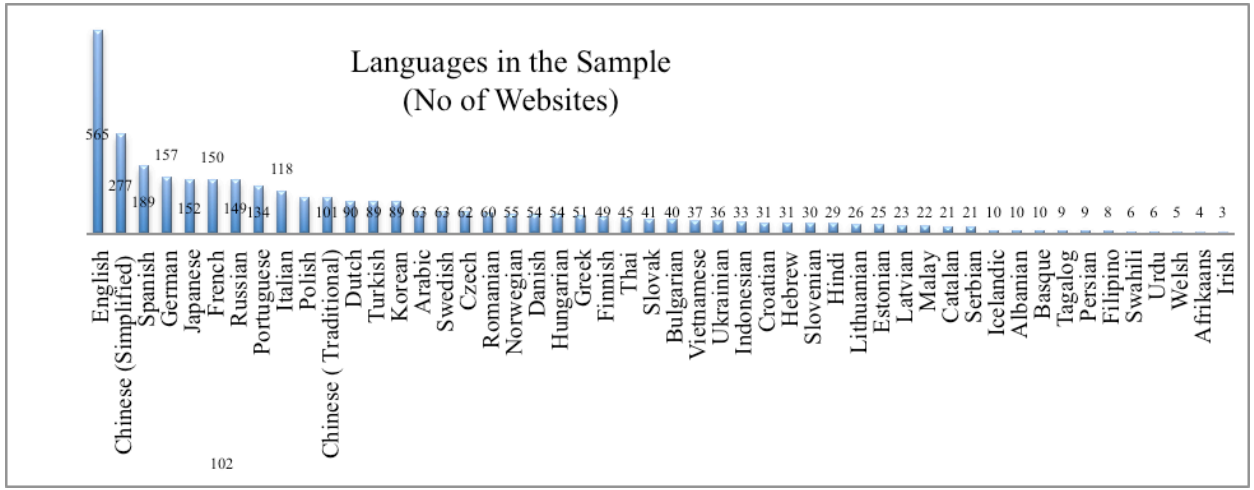

*Figure 2: Countries in the Sample.*

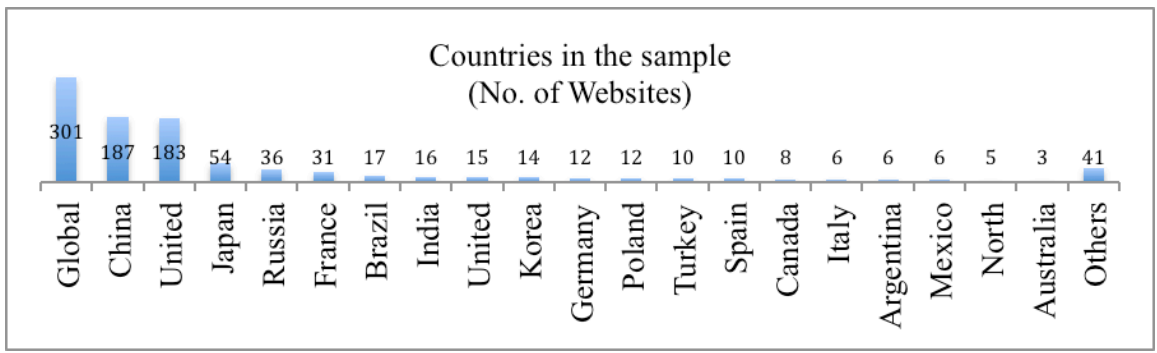

**Figure 3:** *Visualization of the Audience Network.*

*Dots indicate nodes (Web domains). The colors indicate the focal geography of the Web domain.*

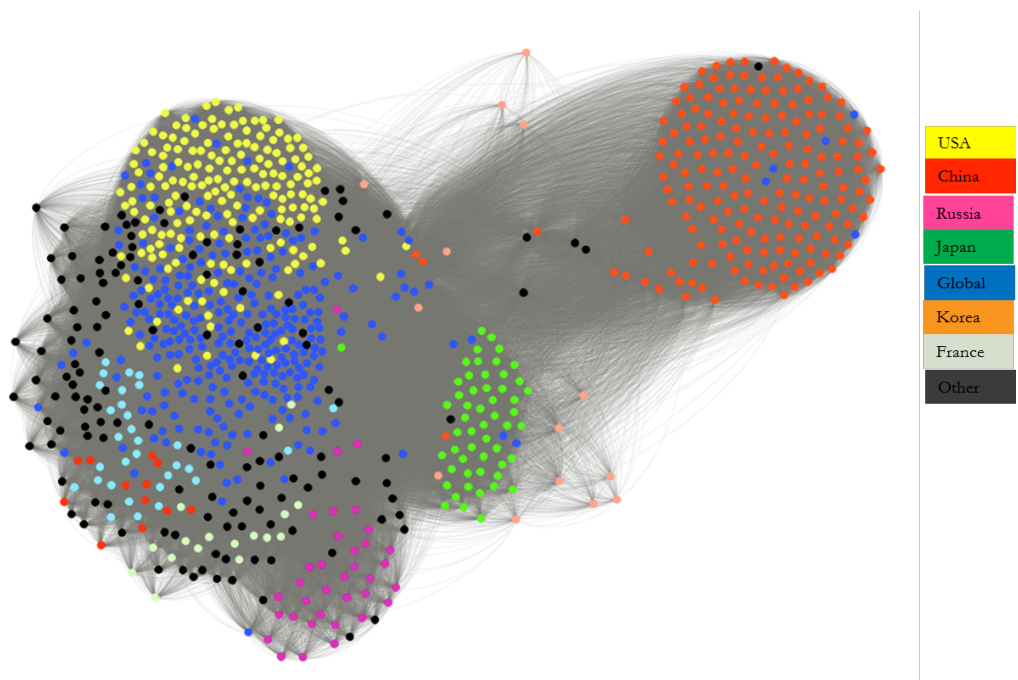

**Figure 4:** *Interaction of Language and Geography on Audience Duplication*

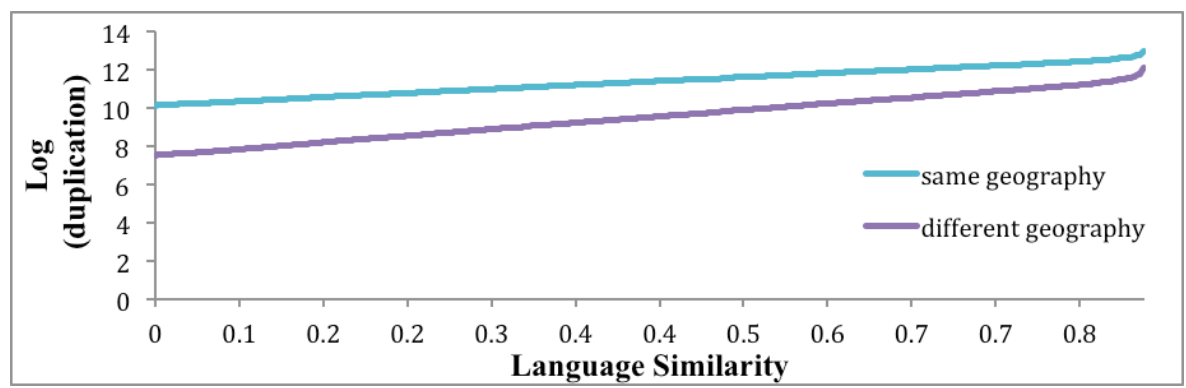